\documentclass[12pt]{article}
\usepackage{a4wide}
\usepackage{graphicx}
\usepackage{axodraw}

\newcommand{\be}{\begin{equation}}
\newcommand{\ee}{\end{equation}}
\newcommand{\ba}{\begin{eqnarray}}
\newcommand{\ea}{\end{eqnarray}}

\begin{document}

\begin{titlepage}
\begin{flushright}
LU TP 09-26\\
September 2009
\end{flushright}
\vfill
\begin{center}
{\Large\bf Leading Logarithms in the Massive $O(N)$ Nonlinear Sigma Model}
\vfill
{\bf Johan Bijnens and Lisa Carloni}\\[0.3cm]
{Department of Theoretical Physics, Lund University,\\
S\"olvegatan 14A, SE 223-62 Lund, Sweden}
\end{center}
\vfill
\begin{abstract}
We review  B\"uchler and Colangelo's result
that leading divergences
at any loop order can be calculated using only one-loop calculations
and we provide an alternative proof.
We then use this method to calculate the leading divergences of
and thus the leading logarithmic corrections to the meson mass in
the massive $O(N)$
nonlinear sigma model to five-loop order.
We also calculate the all-loop result to leading order in the large $N$
expansion by showing that only cactus diagrams contribute and by summing
these via a generalized gap equation.
\end{abstract}
\vfill
\vfill
\end{titlepage}

\section{Introduction}

In a renormalizable theory the use of Renormalization Group Equations (RGE) is
common practice.
RGEs have not yet received
the same attention in non renormalizable effective theories. This is partially
due to the fact that one does not normally have the problem of evolving coupling
constants through a large energy range, and partially to the fact that RGEs in
non renormalizable theories get more complicated as one goes to higher orders.
Loop corrections however can be significant \cite{Weinberg1,GL1,GL2} and need
to be dealt with. 

Consider for example
the scattering length $a^0_0$  in $\pi \pi$ $S$-wave,
$I=0$ scattering in ChPT. Close to threshold, this amplitude may be
expressed in terms of the expansion parameter $(M_{\pi}/4\pi F_{\pi})^2\sim
0.01$. Despite the smallness of the expansion parameter, the one loop
contributions cause a $28\%$ corrections to the tree level prediction
\cite{GL0}. 
The reason for this is that beyond tree level the
expressions for observables contain non-analytic functions \cite{Li-Pagels}
such as $M^2 \log M^2$ which may be large if $M^2$ is small,
even near threshold. It is only natural then to wonder about the size of 
higher order $n$ contributions like $(M^2 \log M^2)^n$, the so called leading
logarithms, and about the size of their coefficients.

In a renormalizable theory these coefficients are fully determined by a one-loop
calculation. This is a consequence of the RGEs The difference between a
renormalizable
and a non-renormalizable theory  is that in the first case the counterterms
needed at any given order have the same form, while in the latter case new
ones are needed at every order. Nevertheless,
one can still make predictions on the leading logarithms.
This fact was first pointed out by
Weinberg \cite{Weinberg1} in the context of Chiral Perturbation Theory
(ChPT). He showed that at two loops
the coefficient of leading logarithm $(M^2 \log M^2)^2$ can be
determined simply by performing a one-loop calculation. 

This method has since been used in ChPT \cite{Weinberg1,GL1,GL2}
to two loop order in
$\pi\pi$ scattering \cite{Colangelo1} and in general \cite{BCE0}. 
It is no longer much used in the purely mesonic sector,  since
most processes are actually fully known at two-loop order as reviewed
in e.g. \cite{Bijnens1}.

In the last few years though, Weinberg's argument has received new attention
\cite{BC,BF1,BF2,Kivel1,Kivel2}, especially
since B\"uchler and Colangelo were able to generalize the result
to all orders \cite{BC}. They showed explicitly that one can obtain the
 (coefficient of the) leading logarithm
at any order by simply performing one-loop calculations and that this
coefficient is just a function of the lowest order coupling
constants. The relevant part of their paper \cite{BC} and their algorithm to
find the coefficients is described in Sect.~\ref{RGE}.

In Section~\ref{RGEalternative} we provide an alternative proof of their
results which does not rely explicitely on $\beta$-functions but follows
directly from the fact that all non-local divergences must cancel.
This version of the proof has the benefit that it shows immediately that one
only needs to calculate the divergent part without worrying about classifying
higher order Lagrangians and that there is a direct link between 
divergences and leading logarithms.

We then apply this method to the case of a \emph{massive} $O(N)$ non-linear
sigma model and calculate the corrections to the meson mass up to five
loops. Sect.~\ref{calculation} contains a detailed explanation of our
calculation. 
Similar calculations have been performed by \cite{Kivel1,Kivel2} who 
showed how to obtain the leading logarithms in the
massless case by deriving a recursion relation for all possible vertices with
up to four mesons. Since in the massless limit the tadpoles vanish,
this allows
obtain the leading logarithms in a straightforward fashion. 

The authors of \cite{BF1} instead calculated the two-point function up to five
loops in ChPT in the chiral limit using dispersive methods\footnote{In the
same paper the authors also calculate the dispersive part of the
three-loop  pion form factor.}. 
Once the first five leading logarithms where known, the next step was finding an
algorithm that would allow them to calculate the $n$-th order one and
eventually to resum the series. In paper \cite{BF2} they considered
a \emph{linear
sigma} model and compared the correlator leading logarithms
they found with those from
ChPT, both in the chiral limit. They showed that it is not possible to
simply use RGEs in the linear sigma model to resum the chiral logarithm series.
The two scales present in the linear sigma model both generate logarithms
that cannot be disentangled.

We are however able to calculate the the five-loop meson mass in the
\emph{massive non-linear} case. What allows us to
obtain these results is the observation that while one needs a
\emph{complete} Lagrangian to any order, this does not have to be
\emph{minimal} (see Sect.~\ref{nonminimal}),
nor does one need its explicit form in terms of chirally invariant operators.
This, combined with the power of FORM \cite{FORM}, allows us to calculate
the relevant parts to five-loop order for the meson mass.

A calculation of this magnitude needs as many checks as possible to
ensure that no mistakes are made. We use two main checks. We perform the
calculations in different parametrizations of the $O(N)$ nonlinear
sigma model. Since different parametrizations distribute  contributions
very differently over the the various Feynman diagrams, the agreement
 provides a rather stringent check. Another check is that the leading
term in $N$ agrees with the result of the large $N$ expansion of the
model.

To the best of our knowledge, this is the first time a study of
the massive $O(N)$ is
performed in this limit, whereas there is a vast literature on large $N$ in
the massless case, especially for the linear sigma model,
see \cite{CJP} and papers citing it.

We solve the mass in the $O(N)$ model in a fashion similar to that often used
in the Nambu-Jona-Lasinio model, see \cite{ENJLreview} and
references therein.
We first proof that only ``cactus'' diagrams
contribute and that they can all be recursively generated by an equation for the
exact propagator, a generalization of the usual gap equation. 
Finally, we are able to find a compact implicit expression for the
all-loop physical mass.  A thorough explanation of the study can be
found in Section~\ref{largeN}.

This paper is organized as follows. In Sect.~\ref{RGE} we discuss
the results of \cite{BC} (Sect.~\ref{RGEdivergences})
and use their proof to show that one does not need a minimal
nor a fully symmetrically formulated Lagrangian at higher orders
(Sect.~\ref{nonminimal}). Sect.~\ref{RGEalternative}
provides an alternative proof of the results of \cite{BC}.
Sect.~\ref{sigmamodel} defines the $O(N)$ nonlinear sigma model
and the different parametrizations that we use.
Section~\ref{largeN} discusses the large $N$ case to all orders.
The calculation of the leading divergences for the
mass to five-loop order in the general case is described in
Sect.~\ref{calculation},
there we also give the result of the calculation.
Sect.~\ref{conclusions} summarizes our main results.
The way we perform the integrals is described in App.~\ref{integrals}.

\section{Renormalization group arguments}
\label{RGE}

\subsection{The equations for the divergences}
\label{RGEdivergences}

This subsection recapitulates the parts of \cite{BC} we will
use. We use dimensional regularization with $d=4-w$.
The Lagrangians can be ordered in an expansion in $\hbar$. We denote
the lowest order Lagrangian with $\mathcal{L}_0$.
\ba
\label{bare}
\mathcal{L}^\mathrm{bare}
&=&\sum_{n\ge 0}\hbar^n \mathcal{L}_n^\mathrm{bare}\,,
\nonumber\\
 \mathcal{L}_n^\mathrm{bare}&=& 
    \frac{1}{\mu^{nw}} \left( \mathcal{L}_n+\mathcal{L}_n^\mathrm{div}\right)\,.
\ea
The divergent part contains the inverse powers of $w$ needed for the
subtraction of the loop divergences at order $n$  \footnote{ When comparing 
with ChPT one should remember that our order $\hbar^n$ 
corresponds to the order $p^{2n+2}$ in ChPT.}. 

 We now expand each term
into a set of $N_n$ operators $\mathcal{O}^{(n)}_{i}$: 
\ba
\label{div}
\mathcal{L}_n &=& \sum_{i=1}^{N_n} c^{(n)}_{i}\mathcal{O}^{(n)}_{i}\,,
\nonumber\\
\mathcal{L}_n^\mathrm{div} &=&
\sum_{i=1}^{N_n} \left(\sum_{k=1}^n\frac{\mathcal{A}^{(n)}_{ki}} {w^{k}}\right)
  \mathcal{O}^{(n)}_{i}\,.
\ea

One key difference between a renormalizable Lagrangian and these is that here
the minimal basis of operators $\mathcal{O}^{(n)}_i$ grows with $n$.
In the remainder, we will assume that all 
one particle irreducible diagrams (1PI) are
made finite separately. This simplifies the calculations
and arguments. That this can always be done is discussed in \cite{BC}.
We have already used here, in the expression of $\mathcal{L}_n^\mathrm{div}$,
the fact that all divergences are local.
The $c^{(n)}_{i}$ are usually referred to as Low-Energy-Constants (LECs).

The bare Lagrangian is $\mu$ independent. This leads to the equations
\ba
\label{RGE1}
0 &=&\mu\frac{d}{d\mu} \mathcal{L}_n^\mathrm{bare}
\nonumber\\
&=&\frac{1}{\mu^{nw}}\sum_{i=1}^{N_n}\mathcal{O}_i
\left[
-nw c^{(n)}_{i}+\mu\frac{d}{d\mu}c^{(n)}_{i}+\sum_{k=1}^n
\left(-nw +\mu\frac{d}{d\mu}\right)\frac{\mathcal{A}^{(n)}_{ki}}{w^{k}}
\right]\,.
\ea
These must be fulfilled separately for each $n$, $i$ and inverse power of $w$.
We define the $\beta$ functions via
\be
\mu\frac{d}{d\mu}c^{(n)}_{i}=\beta^{(n)}_{i}+nw c^{(n)}_{i}\,.
\ee
The RGEs of (\ref{RGE1}) thus become
\be
\label{RGE2}
\beta^{(n)}_{i}+\sum^n_{k=1}
\left(-nw +\mu\frac{d}{d\mu}\right)\frac{\mathcal{A}^{(n)}_{ki}}{ w^{k}} = 0\,.
\ee
The $\mathcal{A}^{(n)}_{ki}$ do not depend explicitely on the $\mu$. They do,
however,
depend upon the $c^{(n)}_{i}$, since they must cancel the divergences stemming
from the loops of the lower order Lagrangians.  We can thus simplify
(\ref{RGE2}) to 
\be
\label{RGE3}
\beta^{(n)}_{i}+\sum^n_{k=1}\left[-nw+
\sum_{m,j}\left(mw c^{(m)}_{j}+
\beta^{(m)}_{j}\right)\frac{\partial}{\partial c^{(m)}_{j}}
\right]
\frac{\mathcal{A}^{(n)}_{ki}}{ w^{k}} = 0\,.
\ee
The loop contributions must be polynomials in the coupling
constants $c^{(m)}_{i}$. It follows that $\mathcal{A}^{(n)}_{ki}/ w^k$ must
also be a polynomial. This means that we can split (\ref{RGE3}) further
into separate equations. It must be true for each power in $w$, but  also for
each $c^{(n)}_{i}$ monomial, since these are independent parameters and can,
in principle, be varied freely. The coefficient of each separate monomial 
in the $c^{(n)}_{i}$ in (\ref{RGE3}) then must vanish separately. 
It is this extra information that allows us to
obtain all the leading divergences from one-loop
calculations \cite{Weinberg1,BC}. 

The powers of $\hbar$ in any diagram
come from two places, a factor $\hbar^l$ comes from the number of loops $l$
and the remainder from the $\hbar^n c^{(n)}_{i}$ present in the diagrams
when vertices with $\mathcal{O}^{(n)}_{i}$ occur. This shows that each monomial
in the $c^{(n)}_{i}$ will also come from a well specified loop level.

Let us write out the various equations for the first few orders.
To order $\hbar$, we only get the equations
\be
\label{rge1}
\beta^{(1)}_{i}-\mathcal{A}^{(1)}_{1i} =0\,.
\ee
There is no dependence on any of the higher order LECs.
We now introduce a notation for the LEC dependence. 
For both $\beta^{(n)}_{i}$ and $\mathcal{A}^{(n)}_{ki}$ we add a subscript $l$ 
indicating the loop level it came from and an argument indicating its
polynomial dependence on the $c^{(m)}_{j}$. In order to simplify notation in
the following we shall omit the $i$ index, so that now $\beta^{(n)}_{i}$ and
$\mathcal{A}^{(n)}_{ki}(j)$ become $\beta^{(n)}_l(j)$ and
$\mathcal{A}^{(n)}_{lk}$, but one should remember throughout the calculation
that we are always speaking about the component
$\beta^n_i$ or $\mathcal{A}^{(n)}_{lki}$. 
At one loop this only adds a subscript 1  
\ba
\beta^{(1)} &\rightarrow& \beta^{(1)}_{1}\,
\nonumber\\
\mathcal{A}^{(1)}_{1} &\rightarrow& \mathcal{A}^{(1)}_{11}\,.
\ea
At order $\hbar^2$ the $\beta^{(2)}_{1}$ function and
$\mathcal{A}^{(2)}_{11}$ can have a first order dependence on $c^{(1)}_{j_1}$,
we make this explicit
\ba
\label{hbar_2}
\beta^{(2)} &\rightarrow& \beta^{(2)}_{2}+c^{(1)}_{j_1}\beta^{(2)}_{1}(j_1)\,,
\nonumber\\
\mathcal{A}^{(2)}_{2} &\rightarrow& \mathcal{A}^{(2)}_{22}\,,\nonumber\\
\mathcal{A}^{(2)}_{1} &\rightarrow& \mathcal{A}^{(2)}_{21}
 +c^{(1)}_{j_1}\mathcal{A}^{(2)}_{11}(j_1)\,.
\ea
A sum over $j_1$ is implied. The $j_1$ indicates that one should
  consider the minimal set of operators available at order 1. Putting
  (\ref{hbar_2}) in (\ref{RGE3}) gives three conditions on the various
coefficients of the monomials in the $c^{(n)}_{i}$:
\ba
\label{rge2}
\beta^{(2)}_{2} &=& 2 \mathcal{A}^{(2)}_{21}\,,
\nonumber\\
\beta^{(2)}_{1}(j_1) &=& \mathcal{A}^{(2)}_{11}(j_1)\,,
\nonumber\\
2\mathcal{A}^{(2)}_{22} &=& \beta^{(1)}_{1}(j_1)\mathcal{A}^{(2)}_{11}(j_1)\,.
\ea
At order $\hbar^3$ the polynomial dependence gets more complicated 
\ba
\label{hbar3}
\beta^{(3)} &\rightarrow& \beta^{(3)}_{3}+c^{(1)}_{j_1}\beta^{(3)}_{2}(j_1)
+c^{(2)}_{j_2} \beta^{(3)}_{1}(j_2)
+c^{(1)}_{j_1}c^{(1)}_{k_1}\beta^{(3)}_{1}(j_1k_1)\,,
\nonumber\\
\mathcal{A}^{(3)}_{3} &\rightarrow& \mathcal{A}^{(3)}_{33}\,,
\nonumber\\
\mathcal{A}^{(3)}_{2} &\rightarrow& \mathcal{A}^{(3)}_{32}
 +c^{(1)}_{j_1}\mathcal{A}^{(3)}_{21}(j_1)\,,
\nonumber\\
\mathcal{A}^{(3)}_{1} &\rightarrow& \mathcal{A}^{(3)}_{31}
 +c^{(1)}_{j_1}\mathcal{A}^{(3)}_{21}(j_1)
+c^{(2)}_{j_1} \mathcal{A}^{(3)}_{11}(2j)
 +c^{(1)}_{j_1}c^{(1)}_{k_1}\mathcal{A}^{(3)}_{11}(j_1k_1)\,.
\ea
Which terms can show up at which level follows from the $\hbar$ counting and the
fact that $l$ loops can at most diverge like $1/w^l$.
Putting (\ref{hbar3}) in (\ref{RGE3}) gives the relations from the
$\mathcal{O}(w^0)$ in (\ref{RGE3})
\ba
\beta^{(3)}_{3}&=&3\mathcal{A}^{(3)}_{31}\,,
\nonumber\\
\beta^{(3)}_{2}(j_1)&=&2\mathcal{A}^{(3)}_{21}(j_1)\,,
\nonumber\\
\beta^{(3)}_{1}(j_2)&=&\mathcal{A}^{(3)}_{11}(j_2)\,,
\nonumber\\
\beta^{(3)}_{1}(j_1k_1)&=&\mathcal{A}^{(3)}_{11}(j_1k_1)\,.
\label{hbar_3}
\ea
The $-nw$ and $\sum_{m,j} mw c^{(m)}_{j}\partial/\partial c^{(m)}_{j}$ terms
in (\ref{RGE3})
always combine to give exactly the loop level back, see \cite{BC} for the
general proof. This gives the first equation in (\ref{hbar_n}).

The $w^{-1}$ part gives
\ba
3\mathcal{A}^{(3)}_{2i} &=&\beta^{(1)}_{1j_1}\mathcal{A}^{(3)}_{21i}(j_1)
+\beta^{(2)}_{2j_2}\mathcal{A}^{(3)}_{11i}(j_2)\,,
\nonumber\\
2\mathcal{A}^{(3)}_{22i}(j_1) &=& 
 2\beta^{(1)}_{1k_1}\mathcal{A}^{(3)}_{11i}(j_1k_1)
+\beta^{(2)}_{1k_2}(j_1)\mathcal{A}^{(3)}_{11i}(k_2)\,.
\ea
Here we have written out the  operator $\mathcal{O}^{(n)}_{i}$ subscripts 
($i$,$j_2, k_1, k_2$). That is to stress the fact that there is a sum over 
a \emph{different} index $j_1\neq i$. We will omit it in the following. 
In deriving the second equation we have used that
$\mathcal{A}^{(3)}_{11i}(j_1k_1)$ is symmetric in $j_1k_1$ and relabeled some
indices. The final equation comes from the $w^{-2}$ part and reads
\be
3 \mathcal{A}^{(3)}_{33} = \beta^{(1)}_{1j_1}\mathcal{A}^{(3)}_{22}(j_1)\,.
\ee
The set of equations gives
\be
\label{rge3}
6\mathcal{A}^{(3)}_{33} =
\beta^{(1)}_{1j_1}\left[2\beta^{(1)}_{1k_1}\mathcal{A}^{(3)}_{11}(j_1k_1)
     +\beta^{(2)}_{1k_1}(j_1)\mathcal{A}^{(3)}_{11}(k_2)\right]\,.
\ee
As one can see, the leading divergence $\mathcal{A}^{(3)}_{33}$ can be
calculated with purely one-loop calculations.

The argument above can be generalized to all orders, see \cite{BC},
here we only quote the results.
We introduce the notation
\be
\nabla_l = \sum^{\infty}_{m=l}\sum_j \beta^{(m)}_{j}
 \frac{\partial}{\partial c^{(m)}_{j}}
\ee
The general set of equations thus reads
\ba
\label{hbar_n}
\beta^{(n)}_{l} &=& l\mathcal{A}^{(n)}_{l1}\,,\hskip3.2cm l=1,\ldots,n\,,
\nonumber\\
l\mathcal{A}^{(n)}_{lk} &=& \sum^{l-k+1}_{l^\prime=1}
     \nabla_{l^\prime}\mathcal{A}^{(n)}_{l-l^\prime,\, k-1}\,,
\hskip1cm l=k,\ldots,n;~k=2,\ldots,n\,.
\ea
For the leading divergence $\mathcal{A}^{(n)}_{nn}$ this equation reads 
\ba
\label{leadingdiv}
n! \mathcal{A}^{(n)}_{nn} &=& \nabla_1^{n-1}\beta^{(n)}_{1}\,,
\nonumber\\
\beta^{(n)}_{1} &=& \mathcal{A}^{(n)}_{11}\,.
\ea
which is a generalization of (\ref{rge1}), (\ref{rge2}) and (\ref{rge3}). This
is the main result of \cite{BC} we will be using. This is a recursive
relation. The order $n=1$ counterterm $\mathcal{A}^{(1)}_{11}\mathcal{O}^{(1)}$
is fixed by the
requirement that it should cancel the $\mathcal{L}_0$ one loop $1/w$
pole. This is now the order $n=1$ coupling in the Lagrangian. The order $n=2$
coupling $\mathcal{A}^{(2)}_{22}$ is fixed by the requirement that it
cancels the one loop $1/w^2$ pole coming form the  $\mathcal{L}_1$  with the
$\mathcal{A}^{(1)}_{11}$ coupling we fixed in the previous step. And so on, 
the $n$-th order $\mathcal{A}^{(n)}_{nn}$ is fixed by the requirement that it 
cancels the $1/w^n$ divergences. These
can be calculated by considering all one loop diagrams generated by the
$\mathcal{A}^{(n-1)}_{n-1, n-1}\mathcal{O}^{(n-1)}$ ,\ldots, 
$\mathcal{A}^{(1)}_{11}\mathcal{O}^{(1)}$ and  $\mathcal{L}_0$ vertices that can
contribute to order $n$.

The renormalized coupling $c^{(1)}$ now contains a $\log \mu$. This
exactly cancels the $\log \mu$ dependence that comes from the
$\mathcal{L}_0$ loop integral, which has the same coefficient as the
divergence. Analogously, when an observable is calculated up to order $n$ its
expression contains a $(\log\mu)^n$ term whose coefficient is given by the
$\mathcal{A}^{(n)}_{nn}$. So (\ref{leadingdiv}) gives a recursive
expression for the coefficients of the leading logarithms.

\subsection{Nonminimal sets of operators}
\label{nonminimal}

In Section \ref{RGEdivergences} we shortly went through the arguments
of \cite{BC} to derive the leading divergence at any order from
only one-loop calculations. In practical applications of the formulas
above one needs a classification of the terms $\mathcal{O}^{(n)}_{i}$ needed
at each order $n$. Determining the complete and minimal set is in general rather
complicated, see e.g. \cite{BCE1} for $\mathcal{L}_2$ in ChPT.
Luckily we do not have to have a minimal and complete Lagrangian in general.

It is sufficient to have a Lagrangian that is complete for the particular
process at hand and lower order Lagrangians that are complete enough so that
all needed $\beta^{(n)}_{i}(1)$ can be obtained. The Lagrangian does not need
to be minimal since the arguments in Section \ref{RGEdivergences} relied
on the fact that all $c^{(n)}_{i}$ can be varied independently.
If we add an irrelevant term, e.g. one that vanishes via partial integration
or other identities, its coefficient can definitely be freely varied
and will not show up in any actual higher order calculations.
This also means that the $\beta$ function of this irrelevant term can be chosen
freely since it will never appear in any expressions.

If we now have two related terms $c^{(n)}_1 \mathcal{O}^{(n)}_1$ and
$c^{(n)}_2 \mathcal{O}^{(n)}_2$, we can always write them as an irrelevant one
$c^{(n)}_{irr} \mathcal{O}^{(n)}_{irr}$
and a relevant one $c^{(n)}_{rel} \mathcal{O}^{(n)}_{rel}$. Since the
irrelevant combination will never appear
and its $\beta^{(n)}_{\mathrm{irr}}$ function is free, we can just as well leave
both terms $c^{(n)}_1 \mathcal{O}^{(n)}_1$ 
$c^{(n)}_2 \mathcal{O}^{(n)}_2$
in the Lagrangian and leave the $\beta^{(n)}_{\mathrm{rel}}$ spread over the
original two terms  $\beta^{(n)}_1$ and
$\beta^{(n)}_2 $. How it is split between the two terms depends
upon the choice of $\beta^{(n)}_{\mathrm{irr}}$, which is free.

In practice, it is sufficient to calculate the divergences
and to express them in some complete set of operators $\mathcal{O}^{(n)}_{i}$.
As long as the set is complete for the given application we will
obtain the correct result.

When constructing a Lagrangian one normally  takes into account all
terms that have the correct symmetry at the required order and then one removes
the so called equation of motion terms, see \cite{BCE1} App.~A, for a
discussion.
In this case, however, we will use neither constraint. We will keep the equation
of motion terms, since then we can make all 1PI
diagrams finite, see the discussion in \cite{BC}. 
We will also use a standard Feynman diagram calculation to obtain 
the infinities and not a more sophisticated method such as e.g. the heat kernel
expansion that was used in \cite{BCE2}. 

The reason is that this way we can use standard
Feynman integral techniques and we do not
have to evaluate all the divergent combinations
of propagators that can appear, see \cite{BCE2,JO},
which is rather difficult at higher orders. The drawback is that this
procedure breaks the symmetries of the
Lagrangian in individual parts of the calculation
even though the final result will respect all symmetry properties
when we use dimensional regularization. However, as we saw in the previous
section, we only use the divergent parts of
these terms and these must obey the symmetries since they are recursively
determined by a symmetric lowest order Lagrangian.

The answer for the divergences for a given process will thus be correct
even without explicitly fixing counterterms with Ward identities.
The correct combinations must show up in our procedure.

The conclusion from this section is that we simply calculate all one-loop
diagrams and rewrite them as terms in the Lagrangians, without bothering
to check if we have a minimal Lagrangian.

We also do not need to have a complete Lagrangian, an operator
$\mathcal{O}^{(n)}_{i}$ will only be relevant if it has a nonzero $\beta$
function. We thus let the calculation itself produce all terms that have a
divergence, give them a coefficient $c^{(n)}_{i}$, and use those in the
equations derived in Section \ref{RGEdivergences}.

For the subleading divergence, the same type of argument shows that it
is sufficient to have the lowest and first order Lagrangian in a symmetric
form to get the subleading divergence at all orders and the obvious
generalization to the further divergences.

\subsection{An alternative proof}
\label{RGEalternative}

A more direct proof of the results of B\"uchler and Colangelo is also possible.

We have presented their method as well since the arguments in
Sect.~\ref{nonminimal} made use of their formulation of the proof in \cite{BC}. 

We rely here on using only 1PI diagrams
and assume they are made fully finite, as was shown to be possible in
\cite{BC} . In this section we obtain the same relations in a more transparent
fashion. We first rewrite (\ref{bare}) and (\ref{div}) as
\be
\mathcal{L}^\mathrm{bare}
 = \sum_{n\ge 0} \hbar^n \mu^{-nw} \sum_i
\left(\sum^n_{k=0}  c^{(n)}_{ki} w^{-k}\right)\mathcal{O}^{(n)}_{i}\,.
\ee
We introduce the notation $\{c\}^n_l$ to indicate
all possible combinations $c^{(m_1)}_{k_1 j_1}c^{(m_2)}_{k_2 j_2}
\ldots c^{(m_r)}_{k_r j_r}$ with $m_i\ge1$, such that $\sum_{i=1,r} m_i = n$ and
$\sum_{i=1,r} k_i = l$.
The $c^{(n)}_{ki}$ with $k\ge1$ have no direct $\mu$ dependence. They only
depend on $\mu$ through their dependence on lower order parameters.
The $c^{(n)}_{0i}$ do depend directly on $\mu$.
Note that since we consider only 1PI diagrams
we have that $\{c\}^n_n = \{c^{(n)}_{ni}\}$.

We denote the contribution from all $l$-loop diagrams at order $\hbar^n$
as $L^n_l$ and we expand this as
\be
L^n_l = \sum^l_{k=0} L^n_{lk} w^{-k}\,.
\ee
This only includes the divergences coming from the loop integrations,
not those from the coefficients in the Lagrangian.

A main observation \cite{BC} is that a given loop level
at a given order $\hbar^n$ always comes with the same power of $\mu$
because of the way the powercounting works.

We can now study the contributions at the different orders in $\hbar$
and $1/w$. For clarity we add here as well which combinations
of couplings of order $n\ge1$ the results depend on.

At order $\hbar^0$ we have only $L^0_0$. At order $\hbar^1$ we have
\be
\label{hbar1}
\frac{1}{w}\left(\mu^{-w} L^1_{00}(\{c\}^1_1)+L^1_{11}\right)
+\mu^{-w} L^1_{00}(\{c\}^1_0)+L^1_{10}\,.
\ee
The divergence must cancel so to get the divergent combinations we have that
\be
\label{hbar1sol}
L^1_{00}(\{c\}^1_1) = -L^1_{11}\,.
\ee
This allows to determine the divergences that need to be subtracted from
a one-loop calculation and it shows that by expanding $\mu^{-w}$
and taking $w\to0$ the explicit $\log\mu$ dependence of any process
is
\be
-\log\mu~  L^1_{00}(\{c\}^1_1) = \log\mu~L^1_{11}\,.
\ee
At order $\hbar^2$ the full contribution is
\ba
\label{hbar2}
&&\frac{1}{w^2}\left(\mu^{-2w} L^2_{00}(\{c\}^2_2)+\mu^{-w}L^2_{11}(\{c\}^1_1)
+L^2_{22}
\right)
\nonumber\\&&
+\frac{1}{w}\left(\mu^{-2w} L^2_{00}(\{c\}^2_1)+\mu^{-w}L^2_{11}(\{c\}^1_0)
+\mu^{-w}L^2_{10}(\{c\}^1_1)+L^2_{21}
\right)
\nonumber\\&&
+\left(\mu^{-2w} L^2_{00}(\{c\}^2_0)+\mu^{-w}L^2_{10}(\{c\}^1_0)
+L^2_{20}
\right)\,.
\ea
All divergences must cancel, also those with powers of $\log\mu$.
If we only look at the parts with $1/w^2$ and $\log\mu/w$
we obtain two equations
\ba
&& L^2_{00}(\{c\}^2_2) + L^2_{11}(\{c\}^1_1) +L^2_{22} = 0\,,
\nonumber\\&&
2 L^2_{00}(\{c\}^2_2) + L^2_{11}(\{c\}^1_1) = 0
\ea
The difference in the coefficients from the first to the second equation
comes from the expansion of the different powers of $\mu^{-w}$.
These equations have the solution
\ba
\label{hbar2sol}
 L^2_{00}(\{c\}^2_2) &=& L^2_{22}\,,
\nonumber\\
 L^2_{11}(\{c\}^1_1) &=& -2 L^2_{22}\,.
\ea
The leading logarithm can be obtained by expanding $\mu^{-w}$
in (\ref{hbar2}) and using (\ref{hbar2sol}):
\be
\frac{1}{2}\log^2\mu\left(4 L^2_{00}(\{c\}^2_2)L^2 +  L^2_{11}(\{c\}^1_1)
\right) =
\log^2\mu~L^2_{22}\,.
\ee
So here we reproduce the known result and that it can be obtained from
a one-loop calculation.
The calculation for the next two orders follows the same lines. A clear pattern emerges. 

At order $\hbar^n$,
the leading part is given by
\be
\frac{1}{w^n}\left(
\mu^{-nw}L^n_{00}(\{c\}^n_n)
+\mu^{-(n-1)w}L^n_{11}(\{c\}^{n-1}_{n-1})
+\cdots
+\mu^{-w} L^n_{n-1~n-1}(\{c\}^{1}_{1})
+L^n_{nn}
\right)\,.
\ee
This part is the one that contributes to the
$1/w^n,\log\mu/w^{n-1},\ldots,\log^{n-1}\mu/w$ divergences
leading to the set of equations: 
\be
\label{hbarn}
\sum^n_{i=0} i^j L^n_{n-i~n-i}(\{c\}^{i}_{i}) = 0\hskip3.0cm j=0,..,n-1 .
\ee
with $0^0=1$ and $L^n_{nn}(\{c\}^{0}_{0}) = L^n_{nn}$.
The generalization of the solution then is
\be
\label{hbarnsol}
L^n_{n-i~n-i}(\{c\}^{i}_{i}) = 
(-1)^i\left(\begin{array}{c}n\\i\end{array}\right) L^n_{nn}\,.
\ee
We can prove that this solves the equations (\ref{hbarn})
above by observing that they can be written as
\be
\lim_{a\to0}\left(\sum^n_{i=0} \left(a\frac{d}{da}\right)^j
 a^{n-j} L^n_{jj}(\{c\}^{n-i}_{n-i})\right) = 0\,.
\ee
Plugging in (\ref{hbarnsol}) we see that this becomes
\be
\lim_{a\to 1}  \left(a\frac{d}{da}\right)^j (-a+1)^n L^n_{nn} = 0\,,
\ee
which is clearly satisfied.
Using $\lim_{a\to 1}\left(a\frac{d}{da}\right)^n (-a+1)^n = (-1)^n n!$
one can also derive that the dependence on $\log^n\mu$ is
\be
\log^n\mu~L^n_{nn}\,.
\ee
This completes our alternative proof of the main result of \cite{BC}.

\section{The $O(N)$ nonlinear sigma model}
\label{sigmamodel}

The $O(N+1)/O(N)$ nonlinear sigma model has as Lagrangian
\be
\label{sigmalag}
\mathcal{L}_{n\sigma} = \frac{F^2}{2}
\partial_\mu \Phi^T\partial^\mu \Phi+F^2 \chi^T \Phi\,. 
\ee
$\Phi$ is a real $N+1$ vector that transforms as the fundamental
representation of $O(N+1)$ and satisfies the constraint $\Phi^T\Phi=1$.
The second term is the one that breaks the symmetry explicitly
by setting
\be
\chi^T = \left(M^2~0 \ldots 0\right)\,.
\ee
The vacuum is given by
\be
\label{vacuum}
\langle \Phi^T \rangle = \left(1~0 \ldots 0\right)\,,
\ee
which breaks the $O(N+1)$ spontaneously to $O(N)$. There is both a spontaneous
symmetry breaking triggered by the vacuum (\ref{vacuum}) and an explicit one
given by $F^2 \chi^T \Phi$. 

This Lagrangian corresponds to the lowest order Lagrangian
of two-(quark-)flavour Chiral Perturbation Theory
for $N=3$ \cite{Weinberglag,GL1}
and has been used to describe alternative Higgs sectors in several
beyond the Standard Model scenarios.

As mentioned in the introduction, we make use of different parametrizations to
check the validity of our results. We write $\Phi$
in terms of a real
$N$-component \footnote{We refer to these as a flavour components.} vector
$\phi$, which transforms linearly under the unbroken part
of the symmetry group, $O(N)$. We use here four different ways
to do this parametrization
\ba
\label{parGL}
\Phi_1 &=& \left(\begin{array}{c}\sqrt{1-\frac{\phi^T\phi}{F^2}}\\
\frac{\phi^1}{F}\\\vdots
\\\frac{\phi^N}{F}\end{array}\right) = 
\left(\begin{array}{c}\sqrt{1-\frac{\phi^T\phi}{F^2}}\\
\frac{\phi}{F}\end{array}\right)
\,,
\\
\label{parWE}
\Phi_2 &=& \frac{1}{\sqrt{1+\frac{\phi^T\phi}{F^2}}}
\left(\begin{array}{c}1\\\frac{\phi}{F}\end{array}\right)
\,,
\\
\label{par3}
\Phi_3 &=&
\left(\begin{array}{c}1-\frac{1}{2}\frac{\phi^T\phi}{F^2}
\\[2mm]
\sqrt{1-\frac{1}{4}\frac{\phi^T\phi}{F^2}}\frac{\phi}{F}\end{array}\right)
\,,
\\
\label{par4}
\Phi_4 &=&
\left(\begin{array}{c}\cos\sqrt{\frac{\phi^T\phi}{F^2}}
\\[2mm]
\sin\sqrt{\frac{\phi^T\phi}{F^2}}\,\frac{\phi}{\sqrt{\phi^T\phi}}\end{array}\right)
\,.
\ea
$\Phi_1$ is the parametrization used in \cite{GL1}, $\Phi_2$
the one originally introduced by Weinberg \cite{Weinberglag}.
$\Phi_3$ is such that the explicit symmetry breaking term in (\ref{sigmalag})
only gives a mass term to the $\phi$ field but no vertices.
$\Phi_4$ is the parametrization one ends up with if using the general
prescription of \cite{CWZ}.
These are all examples of the parametrization that keeps the $O(N)$
symmetry manifest:
\be
\Phi=\left(\begin{array}{c}
\sqrt{1-\frac{\phi^T\phi}{F^2}f^2\left(\frac{\phi^T\phi}{F^2}\right)}
\\
f\left(\frac{\phi^T\phi}{F^2}\right)\,\frac{\phi}{F}
\end{array}
\right)\,.
\ee
Here $f(x)$ is any function with $f(0)=1$.

One thing is worth mentioning. In this work we always calculate with the
usual Feynman diagram techniques. In our calculation we split the
$N$-vector field into an external ($\phi_E$) and a loop ($\xi$) field,
$\phi\to \phi_E+\xi$. The divergence structure we obtain is expressed in
terms of $\phi_E$ and we then set $\phi_E\to\phi$ and use that as input
for the next step.
Splitting $\phi$ in this way, the
symmetry is no longer manifest in each term. Since we are renormalizing these
terms, it would be nice if they were obviously symmetric.
We could have used the
background field method, used e.g. in \cite{GL1,BCE2}, and split into a
classical and a quantum field with well defined symmetry properties
and then calculated the divergent part up to a given number of external legs.
This way we would have been assured that our divergent Lagrangian can be
rewritten into terms fully obeying the symmetry. Once that is done, we could
then use the symmetric quantum field again for the next step.
The problem is that the rewriting into symmetric terms is not easy to implement.
We have shown in
Sec.~\ref{nonminimal} that our method gives the correct answer too.

One additional way to check that our results are correct is to compare with
the known results. For $N=3$ our $O(N)$ corresponds to 
$SU(2)\times SU(2)/SU(2)$, and
loop corrections to the pion mass up to two loops are fully known in this model
\cite{BCEGS,Burgi}. The leading terms at two-loop order were first
obtained in \cite{Colangelo1}. They read
\ba
\label{2loop}
M^2_{phys} &=& M^2\left(1-\frac{1}{2}L_M +\frac{17}{8} L_M^2+\cdots\right)\,,
\nonumber\\
L_M &=& \frac{M^2}{16\pi^2 F^2}\log\frac{ \mu^2}{\mathcal{M}^2}\,.
\ea
where in numerical applications one usually chooses $\mathcal{M}=M$.

\section{The large $N$ approximation}
\label{largeN}

The linear sigma model has been treated very much in the large $N$
approximations. The literature can be traced back to \cite{CJP} but
the literature on the nonlinear sigma model is smaller. In addition,
it is mainly restricted to the massless case while here we are interested
in the massive case. \cite{DobadoPelaez} did include masses
but only to first order.
There are some subtleties involved in large $N$ in
effective theories because of the presence of the higher order
Lagrangians, see e.g. \cite{WeinberglargeN}. We however keep our discussion
on the level of the loop diagrams with the lowest order Lagrangian and
stick to $L^n_{nn}$ in the notation of Sect.~\ref{RGEalternative}

\begin{figure}
\begin{center}
\setlength{\unitlength}{1pt}
\begin{picture}(115,150)
\SetScale{1.0}
\Line(25,0)(50,20)
\Line(75,0)(50,20)
\Oval(50,45)(25,15)(0)
\Oval(50,95)(25,15)(0)
\Oval(50,135)(15,7)(0)
\Oval(50,105)(15,7)(0)
\Oval(90,45)(25,15)(90)
\Oval(90,75)(15,7)(0)
\Oval(90,15)(15,7)(0)
\Oval(17.5,95)(17.5,8)(90)
\Oval(17.5,120)(17,8)(0)
\Oval(17.5,70)(17,10)(0)
\Oval(17.5,79)(8,5)(0)
\Oval(17.5,92)(5,5)(0)
\Oval(70,95)(5,5)(0)
\end{picture}
\end{center}
\caption{\label{figcactus} A typical diagram that contributes at leading
order in $N$. Note that vertices can have many different loops attached
since the Lagrangians contain vertices with any number of fields.
The flavour-loops coincide with the loops in momentum.}
\end{figure}
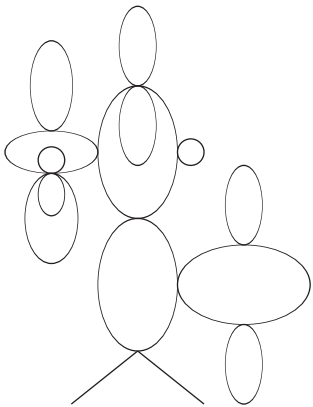
We choose here the Lagrangian to be extensive in $N$. This means we have
to choose $F^2\propto N$. Looking at the Lagrangians in Sect.~\ref{sigmamodel}
we easily see that vertices with $2n$ legs have a factor
$F^{2-2n}$ and are thus suppressed by $N^{1-n}$.
Extra factors of $N$ come from closed loops in the flavour index.
It is sufficient to look at one-particle-irreducible (1PI) diagrams, the flavour
indices for lines not inside a loop are determined by the external
flavour indices.

We thus look at the one-particle-irreducible diagrams only
and use methods similar to those used in \cite{Weinberg1} for
proving the powercounting.
A given diagram has $N_L$ loops, $N_{2n}$ vertices with $2n$ legs,
$N_I$ (internal) propagators and
$N_E$ external legs. These are related via
\ba
N_L &=& N_I-\sum_n N_{2n}+1\,,
\nonumber\\
2 N_I +N_E&=& \sum_n 2n N_{2n}\,.
\ea
We thus get
\be
N_L = \sum_n (n-1) N_{2n}-\frac{1}{2}N_E+1\,.
\ee
The tree level one-particle-irreducible diagram has one vertex with
$2n=N_E$ and thus comes
with a factor of $N^{1-N_E/2}$.
A generic one-particle-irreducible loop diagram has thus a suppression factor
\be
\label{Ncounting}
N^{-\sum_n (n-1) N_{2n}} = N^{-N_L-N_E/2+1}\,,
\ee
from the vertices. Extra factors of $N$ come from the closed flavour loops
where each closed flavour loop give a factor of $N$. (\ref{Ncounting})
shows that for a diagram to be leading in $N$, 
there must be as many closed
flavour loops as there are loops. Since the $\phi$ fields only carry one flavour index,
this means that each loop must coincide with the flavour loop and there
can be no lines shared between two loops. All diagrams that contribute
to a given process at leading order in $N$ are thus diagrams that
only contain products of one-loop diagrams, these we call
cactus diagrams after the looks of the ``prickly pear'' cactus.
A typical example is shown in Fig.~\ref{figcactus}.

How can we now resum all of these diagrams? The easiest way is
to notice that they can be generated recursively. First we note that
the inverse of the full propagator is given by the inverse of the lowest
order propagator and the sum of all the one-particle-irreducible
diagrams with two external legs. This leads to the equation graphically
depicted in Fig.~\ref{figgap}.
The difference with the usual gap equation in NJL-type theories
as discussed in e.g. \cite{ENJLreview} is that we have an infinite
number of terms here. This actually turns out to be manageable.
\begin{figure}
\begin{center}
\setlength{\unitlength}{1pt}
\begin{picture}(450,90)
\SetScale{1.0}
\SetWidth{3.0}
\Text(0,0)[lb]{\large\boldmath$($}
\Line(5,7)(25,7)
\Text(25,0)[lb]{\large\boldmath$)^{-1}$}
\Text(60,7)[]{\large\boldmath$=$}
\SetWidth{1.0}
\Text(75,0)[lb]{\large\boldmath$($}
\Line(80,7)(105,7)
\Text(105,0)[lb]{\large\boldmath$)^{-1}$}
\Text(140,7)[]{\large\boldmath$+$}
\Line(150,7)(200,7)
\Text(210,7)[]{\large\boldmath$+$}
\Line(220,7)(270,7)
\Text(280,7)[]{\large\boldmath$+$}
\Line(290,7)(340,7)
\Text(350,7)[]{\large\boldmath$+$}
\Line(360,7)(410,7)
\Text(415,7)[l]{\large\boldmath$+\cdots$}
\SetWidth{3}
\Oval(175,17)(10,5)(0)
\Oval(245,17)(10,5)(0)
\Oval(245,22)(15,10)(0)
\Oval(315,17)(10,5)(0)
\Oval(315,22)(15,10)(0)
\Oval(315,27)(20,15)(0)
\Oval(385,17)(10,5)(0)
\Oval(385,22)(15,10)(0)
\Oval(385,27)(20,15)(0)
\Oval(385,32)(25,20)(0)
\end{picture}
\end{center}
\caption{\label{figgap} The graphical representation of the equation
that generates all the cactus diagrams for the propagator.
A thick line indicates the full propagator, a thin line the inverse one.}
\end{figure}

Let us look at the Lagrangians of Sect.~\ref{sigmamodel}.
They all contain at most two derivatives. The loops in Fig.~\ref{figgap}
are all tadpoles and thus produce no extra dependence
on the external momentum $p$. All the dependence
on $p$ must come from the derivatives
present in the vertices.
The full inverse
propagator is thus of the form $Z_P p^2-B_P$, where neither $Z_P$ nor $B_P$
depend on the momentum $p$. This is true for all parametrizations.

We now use the first parametrization. The vertices with derivatives
are generated by
\be
\frac{1}{2 F^2}\frac{1}{1-\frac{\phi^a \phi^a}{F^2}}
\phi^b\partial_\mu\phi^b \phi^c\partial^\mu\phi^c\,,
\ee
where we have written $\phi^T\phi=\phi^a\phi^a$ to bring out the sum over
flavour indices explicitly. Each loop must allow for a sum over the flavour
indices to be leading in $N$. The derivatives must either both act on the
external fields
or both on the same loop
to give a nonzero result\footnote{This will not be true for more
complicated processes but can be dealt with in that case as well\cite{BCwork}.}.
When they act inside a loop,
the fields  $\partial_\mu \phi^b \partial^\mu \phi^c$
must be contracted to form the loop. 
This requires $b=c$ and the flavour in this loop is thus determined
by the outer fields and cannot be separately summed over. 
Consequently, the diagram gives no leading $N$ contribution.

If the derivatives hit the external legs,
the indices in at least one loop are fixed by the external ones and
again cannot be summed over. It follows that the contribution is not leading order in $N$.

In either case the loop diagrams generated by the kinetic term give no leading
$N$ correction, so $Z_P=1$.

Thus we only need to look at the vertices coming
from the mass term
\be
\label{massvertex}
\mathcal{L}_{\mathrm mass}  = F^2 M^2\sqrt{1-\phi^a\phi^a/F^2} \equiv F^2 M^2 f(x)
\equiv F^2 M^2\sum_i a_i x^i\,,
\ee
with $x=\phi^a\phi^a/F^2$.
Again, consider the a loop diagram. The external legs need to come from the same
flavour index otherwise it will not be leading in $1/N$. For each term in 
(\ref{massvertex}) there are $i$ ways to choose which $x$ corresponds to
the external legs
\be
\label{massvertex2}
M^2 \phi_{ext}^a\phi_{ext}^a \sum_{i\ge2} a_i i x^{i-1}
= M^2\phi_{ext}^a\phi_{ext}^a \left(\frac{df}{dx}(x)-a_1\right)\,. 
\ee
The sum in (\ref{massvertex2}) starts from $i=2$ since the $i=0,1$ terms
are a constant and the tree level mass term respectively.
Eq (\ref{massvertex}) then reads
\be
-\frac{1}{2}M^2 \phi_{ext}^a\phi_{ext}^a
\left(\frac{1}{\sqrt{1-\frac{\phi^c\phi^c}{F^2}}}-1\right)\,.
\ee
The leading contribution comes from contracting the fields with the same flavour index.
There is only one way to do this for each term.
Each contraction corresponds to a tadpole in Fig.~\ref{figgap}.
The full result can be written as
\be
p^2-B_P = p^2-M^2-M^2
\left(\frac{1}{\sqrt{1+\frac{N}{F^2}A(B_P)}}-1\right)\,,
\ee
where $iA(B_P)=\int d^d p\,1/(p^2-B_P)$ is the relevant one-loop
tadpole integral.
The all-loop result at leading order in $N$ is thus the solution
of
\be
\label{resultlargeN}
M^2 = M^2_{phys} \sqrt{1+\frac{N}{F^2}A(M^2_{phys})}\,.
\ee
Here $B_P$ coincides with the physical mass squared, $M^2_{phys}$,
since $Z_P=1$.

The same result can be derived in the other parametrizations.
If we take the third one, where the only vertices come from
the term with derivatives, the same type of argument as above
with derivatives
and flavour indices shows
that the only relevant vertex is
\be
-\frac{1}{8F^2}\phi^a\phi^a\partial_\mu\phi^b\partial^\mu\phi^b\,.
\ee
So here the gap equation reduces to the first nontrivial term on the
right-hand-side only. The structure of the inverse propagator
is still $Z_P p^2-B_P$ and the gap equation leads to two
equations with $M^2_{phys}=B_P/Z_P$
\ba
Z_P &=& 1+\frac{N}{4F^2 Z_P}A(M^2_{phys}),,
\nonumber\\
Z_P M^2_{phys} &=& M^2-\frac{N}{4F^2 Z_P}M^2_{phys} A(M^2_{phys})\,.
\ea
Here we have expressed the integral containing an extra $q^2$
using  (\ref{reduceintegrals}) in terms of the one without.
Solving leads to the solutions
\ba
Z_P &=&\frac{1}{2}\left(1+\frac{M^2}{M^2_{phys}}\right)\,,
\nonumber\\
M^4 &=& M^4_{phys}\left(1+\frac{N}{F^2}A(M^2_{phys})\right)\,,
\ea
which agrees with the previous result (\ref{resultlargeN}).

The leading logarithm can be expressed by replacing
$A(M^2_{phys})$ by
\be
\overline A(M^2_{phys}) = \frac{M^2_{phys}}{16\pi^2}
\mathrm{log}\frac{\mu^2}{M^2_{phys}}\,.
\ee
In terms of
\be
y = \frac{N M^2}{16\pi^2 F^2}\log\frac{\mu^2}{M^2}
\ee
we can invert the result (\ref{resultlargeN}):
\be
\label{resultlargeN2}
\frac{M^2_{phys}}{M^2} =
 1 - \frac{1}{2} y + \frac{5}{8} y^2 - y^3 
 + \frac{231}{128} y^4 - \frac{7}{2} y^5 + \frac{7293}{1024} 
         y^6 - 15 y^7 + \frac{1062347}{32768} y^8 
+ \cdots\,.
\ee
Note that (\ref{resultlargeN}) actually converges faster.
Expanding the square root in
\be
z = \frac{N M^2_{phys}}{16\pi^2 F^2}\log\frac{\mu^2}{M^2_{phys}}
\ee
we have
\be
\label{resultlargeN3}
\frac{M^2}{M^2_{phys}} =
 1 + \frac{1}{2} z - \frac{1}{8} z^2 +\frac{1}{16} z^3 
 - \frac{5}{128} z^4 + \frac{7}{256} z^5 - \frac{21}{1024} z^6
 + \frac{33}{2048} z^7 - \frac{429}{32768} z^8 
+ \cdots\,,
\ee
which has much smaller coefficients than (\ref{resultlargeN2}).

\section{The calculation}
\label{calculation}

\begin{figure}
\setlength{\unitlength}{1pt}
\begin{picture}(50,70)
\SetScale{1}
\SetPFont{Helvetica}{7}
\Oval(25,40)(20,20)(0)
\Line(0,20)(50,20)
\BText(25,20){0}
\Text(25,0)[b]{(a)}
\end{picture}
~
\setlength{\unitlength}{1pt}
\begin{picture}(110,70)
\SetScale{1}
\SetPFont{Helvetica}{7}
\Oval(25,40)(20,20)(0)
\Line(0,20)(50,20)
\BText(25,20){1}
\Oval(85,40)(20,20)(0)
\Line(60,20)(110,20)
\BText(85,20){0}
\BText(85,60){1}
\Text(55,0)[b]{(b)}
\end{picture}
~
\setlength{\unitlength}{1pt}
\begin{picture}(230,70)
\SetScale{1}
\SetPFont{Helvetica}{7}
\Oval(25,40)(20,20)(0)
\Line(0,20)(50,20)
\BText(25,20){2}
\Oval(85,40)(20,20)(0)
\Line(60,20)(110,20)
\BText(85,20){0}
\BText(85,60){2}
\Oval(145,40)(20,20)(0)
\Line(120,20)(170,20)
\BText(145,20){1}
\BText(145,60){1}
\Oval(205,40)(20,20)(0)
\Line(180,20)(230,20)
\BText(205,20){0}
\BText(220,50){1}
\BText(190,50){1}
\Text(115,0)[b]{(c)}
\end{picture}
\caption{\label{fig1PImass} The diagrams needed
up to order $3$ for the inverse propagator.
Vertices of order $\hbar^i$ are indicated with \fbox{i}.
(a) The diagram needed at order $\hbar$.
(b) The 2 diagrams needed at order $\hbar^2$.
(c) The 4 diagrams needed at order $\hbar^3$.}
\end{figure}
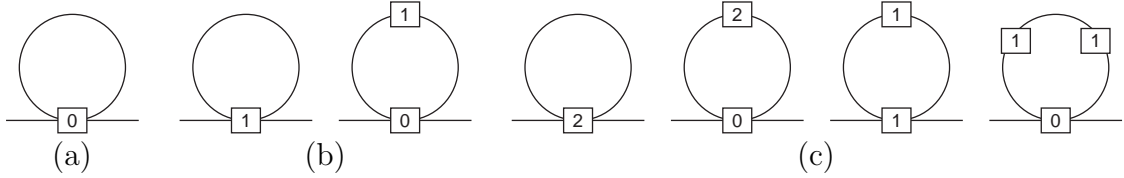
We determine the mass by finding the zero of the inverse propagator.
We therefore must calculate all the one-loop diagrams that are
needed to obtain the divergence of the inverse propagator
to the order desired.
At order $\hbar$ there is only one diagram, at $\hbar^2$
there are 2 and at order $\hbar^3$ there are 4. These
are shown in Fig.~\ref{fig1PImass}.
We have not shown them but at order $\hbar^4$ there
are 7 and at $\hbar^5$ there are 13 diagrams to be calculated.

\begin{figure}
\begin{center}
\setlength{\unitlength}{1pt}
\begin{picture}(110,70)
\SetScale{1}
\SetPFont{Helvetica}{7}
\Oval(25,40)(20,20)(0)
\Line(0,10)(25,20)
\Line(0,25)(25,20)
\Line(50,10)(25,20)
\Line(50,25)(25,20)
\BText(25,20){0}
\Oval(85,40)(20,20)(0)
\Line(60,20)(110,20)
\Line(60,60)(110,60)
\BText(85,20){0}
\BText(85,60){0}
\Text(55,0)[b]{(a)}
\end{picture}
\qquad
\setlength{\unitlength}{1pt}
\begin{picture}(230,70)
\SetScale{1}
\SetPFont{Helvetica}{7}
\Oval(25,40)(20,20)(0)
\Line(0,10)(25,20)
\Line(0,25)(25,20)
\Line(50,10)(25,20)
\Line(50,25)(25,20)
\BText(25,20){1}
\Oval(85,40)(20,20)(0)
\Line(60,10)(85,20)
\Line(60,25)(85,20)
\Line(110,10)(85,20)
\Line(110,25)(85,20)
\BText(85,20){0}
\BText(85,60){1}
\Oval(145,40)(20,20)(0)
\Line(120,20)(170,20)
\Line(120,60)(170,60)
\BText(145,20){1}
\BText(145,60){0}
\Oval(205,40)(20,20)(0)
\Line(180,20)(230,20)
\Line(180,60)(230,60)
\BText(205,20){0}
\BText(205,60){0}
\BText(185,40){1}
\Text(115,0)[b]{(b)}
\end{picture}
\end{center}
\caption{\label{fig1PIpipi} The diagrams needed for the divergence
of the 4-meson vector. (a) The 2 diagrams to order $\hbar$.
(b) the 4 diagrams to order $\hbar^2$
}
\end{figure}
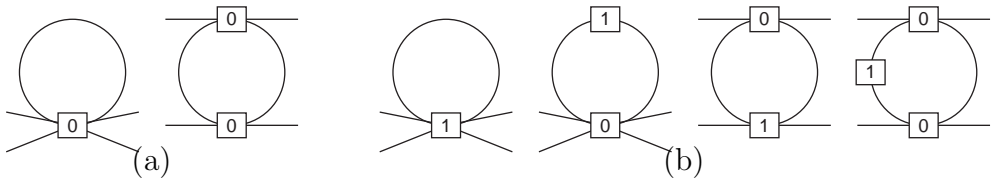

To order $\hbar$ it is sufficient to know the lowest-order Lagrangian,
but at order $\hbar^2$ we need to know the (divergent part of the)
vertices coming from the Lagrangian of order $\hbar$ with two and four external
legs. The diagram of Fig.~\ref{fig1PImass}(a) gives the divergence
of the vertex with two legs but we also need to calculate the
divergence of the vertex with four legs.
This requires the diagrams shown in Fig.~\ref{fig1PIpipi}(a).

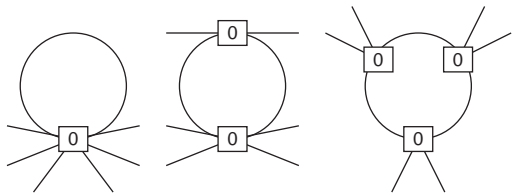
\begin{figure}
\begin{center}
\setlength{\unitlength}{1pt}
\begin{picture}(190,70)
\SetScale{1}
\SetPFont{Helvetica}{7}
\Oval(25,40)(20,20)(0)
\Line(0,10)(25,20)
\Line(0,25)(25,20)
\Line(50,10)(25,20)
\Line(50,25)(25,20)
\Line(10,0)(25,20)
\Line(40,0)(25,20)
\BText(25,20){0}
\Oval(85,40)(20,20)(0)
\Line(60,10)(85,20)
\Line(60,25)(85,20)
\Line(110,10)(85,20)
\Line(110,25)(85,20)
\Line(60,60)(110,60)
\BText(85,20){0}
\BText(85,60){0}
\Oval(155,40)(20,20)(0)
\Line(145,0)(155,20)
\Line(165,0)(155,20)
\Line(130,70)(140,50)
\Line(120,60)(140,50)
\Line(180,70)(170,50)
\Line(190,60)(170,50)
\BText(155,20){0}
\BText(140,50){0}
\BText(170,50){0}
\end{picture}
\end{center}
\caption{\label{fig1PI6pi} The 3 diagrams needed for the divergence
of the 6-meson vector to order $\hbar$.}
\end{figure}

To order $\hbar^3$, we need still more vertices, we need
the divergence of the two-leg vertex to order $\hbar^2$, these
diagrams we already have but we also need the four-leg vertex to
order $\hbar^2$ which can be calculated from the diagrams
in Fig.~\ref{fig1PIpipi}(b). Inspection of the vertices there shows we
already have all we need but for the 6-leg vertex at order $\hbar$.
To obtain that we also need to evaluate all diagrams shown in
Fig.~\ref{fig1PI6pi}.
By now, the pattern should be clear,
to get the mass at order $\hbar^n$ we need
the 2 and four-meson vertex at order $\hbar^{n-1}$,
the 2, 4 and 6-meson vertex at order $\hbar^{n-2}$ and so on.
Similarly one can see that to get the mass at order $\hbar^n$,
we need to calculate one-loop diagrams with up to $n$ vertices.
The extension  to order $\hbar^5$ shows that we
we need to calculate 18, 26, 33, 26 and 13 at orders
 $\hbar^1,\ldots,\hbar^5$ respectively.

We have organized this calculation by first expanding the
lowest-order Lagrangian to the order needed, up to vertices
with 12 mesons for this work.
With these vertices we then calculate all 1PI
diagrams with up to 10 external legs. The divergent part of all
needed integrals can be calculated relatively easily using the
technique described in App.~\ref{integrals}.
At this stage, the dependence on external momenta is also put back as
derivatives on the external legs and everything assembled
to give the divergent part at order $\hbar$ for all the vertices
with up to 10 legs using (\ref{hbar1sol}).
So we have assembled everything we need to calculate
the one-loop divergences to order $\hbar^2$. The 26 diagrams are
evaluated and we obtain the divergences at order $\hbar^2$
using (\ref{hbar2sol}).
The process is then repeated up to order $\hbar^5$.
All of the above steps have been programmed in FORM.
The CPU time needed increases rapidly with the order $n$ one wishes to reach.
The Lagrangians at higher orders tend to contain very many terms
and constructing the diagrams with many external legs at higher orders
is also extremely time consuming.
The CPU time used on a typical PC for the mass-divergence
to order $\hbar^n$
was approximately
0.1 seconds for $\hbar$, 0.3 seconds for $\hbar^2$
11 seconds for $\hbar^3$, 700 seconds for $\hbar^4$
and 30000 seconds for $\hbar^5$.
These running times were achieved after several optimizations in
the choice of routing the external momenta through
the Feynman diagrams.

We have performed the calculation for each of the four parametrizations
shown in Sect.~\ref{sigmamodel}. Since each parametrization distributes
contributions rather differently over the different Feynman diagrams,
this provides a strong check on the consistency of the final result.

We can from these divergences then obtain the leading logarithm.
This leads to the result for the physical mass
\be
\label{mainresult}
M^2_{phys} = M^2\left(1+ a_1 L_M + a_2 L_M^2 +a_3 L_M^3+ a_4 L_M^4 + a_5
L_M^5+\cdots\right)\,,
\ee
where $L_M$ is defined in (\ref{2loop}).
The coefficients $a_1,\ldots,a_5$ are give in Tab.~\ref{tabai}
for $N=3$ and general N.
\begin{table}
\begin{center}
\begin{tabular}{|c|c|l|}
\hline
i & $a_i$ for $N=3$ & $a_i$ for general $N$\\
\hline
\hline
1 & $-1/2$        & $ 1 - 1/2~N$\\
\hline
2 & 17/8       & $7/4
          - 7/4~N
          + 5/8~N^2$\\
\hline
3 & $-103/24$     & $ 37/12
          - 113/24~N
          + 15/4~N^2
          - N^3 $ \\
\hline
4 & 24367/1152 & $ 839/144
          - 1601/144~N
          + 695/48~N^2
          - 135/16~N^3
          + 231/128~N^4 $ \\
\hline
5 & $-8821/144$   & $33661/2400
          - 1151407/43200~N
          + 197587/4320~N^2$\\
& & $
          - 12709/300~N^3
          + 6271/320~N^4
          - 7/2~N^5 $ \\
\hline
\end{tabular}
\end{center}
\caption{\label{tabai} The coefficients $a_i$ of the leading logarithm $L_M^i$
up to $i=5$ for the physical meson mass.}
\end{table}
The result for $N=3$ agrees with the known
results for $a_1$ and $a_2$ given in (\ref{2loop})
and the leading term in $N$ at each order agrees with the
expansion of the all-order result in the large $N$ expansion
(\ref{resultlargeN2}). Note that the large $N$ approximation is definitely not
a good approximation to the $N=3$ coefficients.

The result (\ref{mainresult}) can be inverted and we find
again a better converging expansion.
\be
\label{mainresult2}
M^2 = M^2_{phys}\left(1+ b_1 L_{M_{phys}} + b_2 L_{M_{phys}}^2
 +b_3 L_{M_{phys}}^3+ b_4 L_{M_{phys}}^4 + b_5
L_{M_{phys}}^5+\cdots\right)\,.
\ee
The coefficients $b_1,\ldots,b_5$ are give in Tab.~\ref{tabbi}
for $N=3$ and general N.
\begin{table}
\begin{center}
\begin{tabular}{|c|c|l|}
\hline
i & $b_i$ for $N=3$ & $b_i$ for general $N$\\
\hline
\hline
1 & $1/2$        & $ - 1 + 1/2~N$\\
\hline
2 & $-13/8$       & $1/4
          - 1/4~N
          -1/8~N^2$\\
\hline
3 & $-19/48$     & $ 2/3
          - 11/12~N
          + 1/16~N^3$ \\
\hline
4 & $-5773/1152$ & $ - 8/9 + 107/144~N - 1/6~N^2 - 
         1/16~N^3 - 5/128~N^4$\\
\hline
5 & $ - 3343/768$   & $ - 18383/7200 + 130807/43200~N
          - 2771/2160~N^2 - 527/1600~N^3 $\\
& & $ + 23/640~N^4 + 7/256~N^5$\\
\hline
\end{tabular}
\end{center}
\caption{\label{tabbi} The coefficients $b_i$ of the leading logarithm $L_{M_{phys}}^i$
up to $i=5$ for the lowest order meson mass in terms of the physical mass.}
\end{table}
Just as the coefficients in (\ref{resultlargeN3}) are much smaller
than in (\ref{resultlargeN2}) we see that the $b_i$ are much smaller than
the $a_i$.

\begin{figure}
\begin{minipage}{0.49\textwidth}
\includegraphics[angle=270,width=\textwidth]{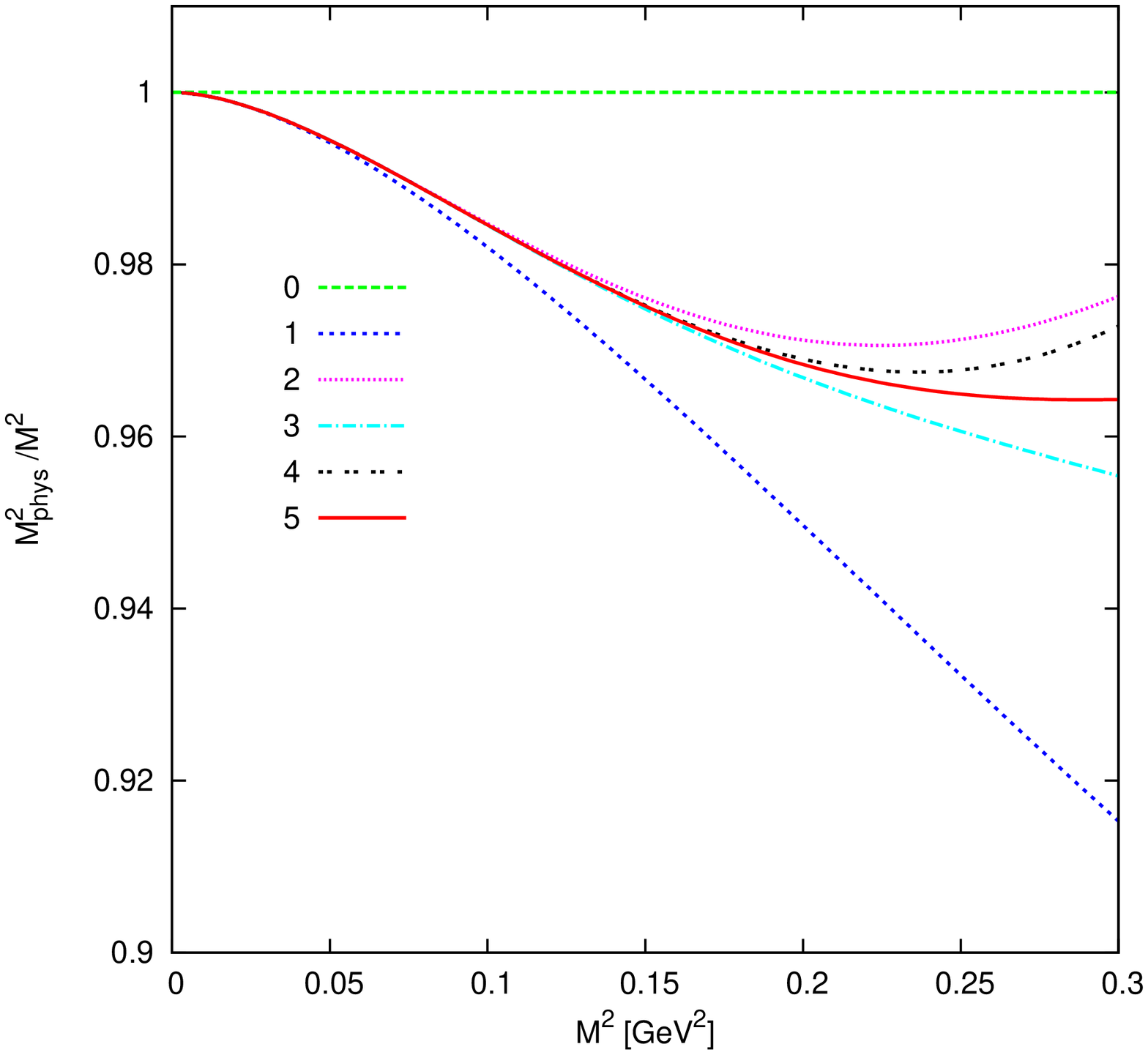}
\centerline{(a)}
\end{minipage}
\begin{minipage}{0.49\textwidth}
\includegraphics[angle=270,width=\textwidth]{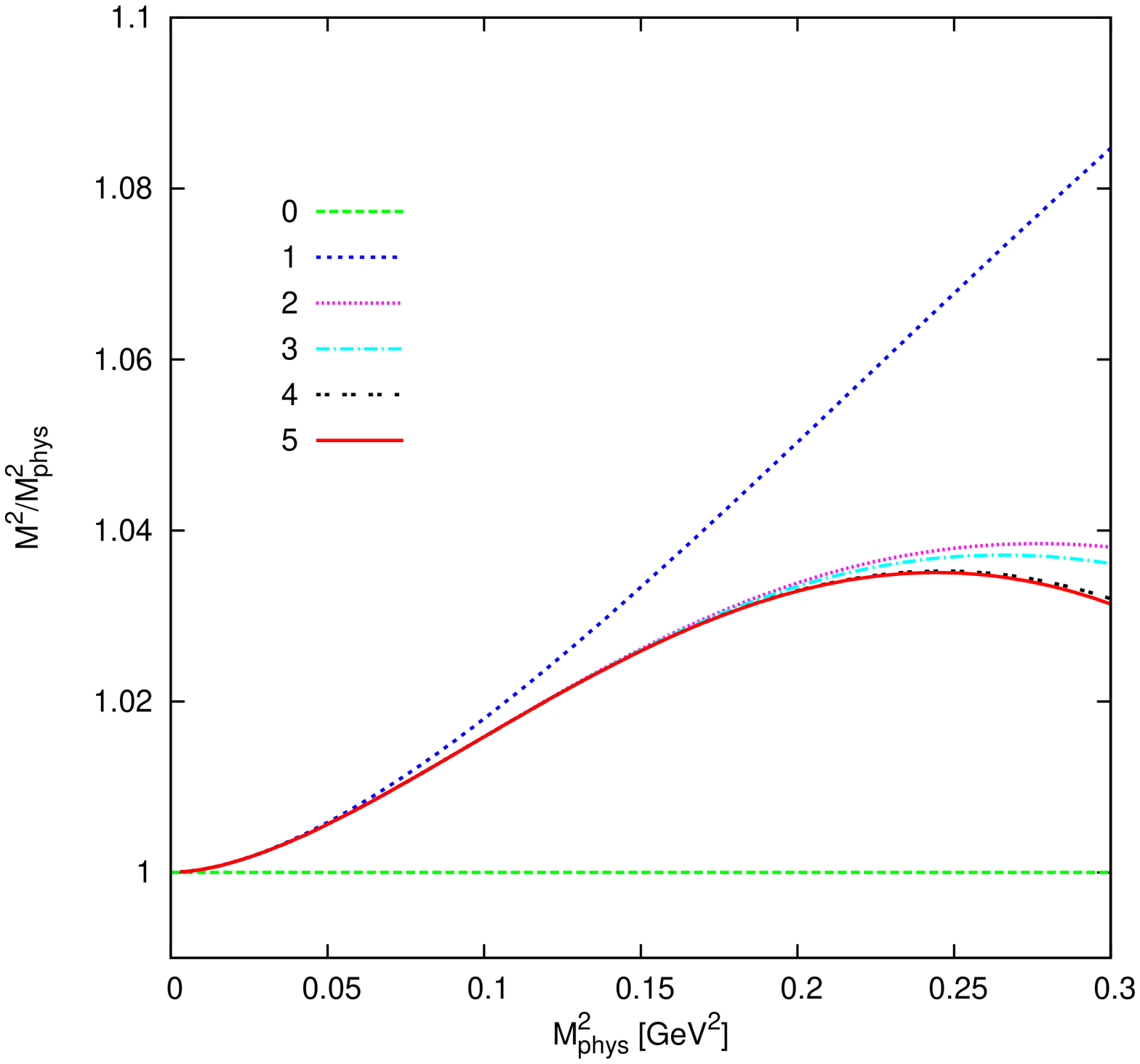}
\centerline{(b)}
\end{minipage}
\caption{\label{figM2}
The expansions of the leading logarithms order by order for $F=0.090$~GeV,
$\mu =1$~GeV and
$N=3$.
(a) $M^2_{phys}$ in terms of $M^2$ Eq.~(\ref{mainresult})
(b) $M^2$ in terms of $M^2_{phys}$ Eq.~(\ref{mainresult2}).}
\end{figure}

In order to get a feeling of the size of these corrections and of the
convergence of the series for the very relevant
case $N=3$ we have plotted them in Fig.~\ref{figM2}. On the left side we see
the result (\ref{mainresult}) and on the right side the result (\ref{mainresult2}) for a value of $F=0.090$~GeV amd $\mu=1$~GeV.
Both clearly converge in the region shown and the inverse
one clearly converges faster.

\section{Conclusions and discussion}
\label{conclusions}

In this paper we have obtained three main results. 

First we provided an
alternative proof for the results of \cite{BC} for the leading divergence
at any loop order. Our proof relies of course on the same physical
principles as the one in \cite{BC} but is simpler
algebraically. 

Our second result is the large $N$ expansion of the
\emph{massive} nonlinear $O(N)$ sigma model, we did not use the analog
of the methods in \cite{CJP} for the massless case but obtained a
recursive equation to sum all the relevant cactus diagrams.
This method is clearly extendable to other processes than the mass
we considered here.
The formula obtained for the mass in the large $N$ limit (\ref{resultlargeN})
is remarkably simple. Its relevance for the real case $N=3$ is not clear, since
the large $N$ result gives a rather poor approximation to the $a_i$
coefficients when $N=3$, as shown in Tab.~\ref{tabai}.

The third result is the actual calculation for general $N$
of the leading logarithm for the meson mass to five-loop order.
This result agrees with the known two-loop result for $N=3$ and with
the derived result for the leading term in $N$.

It is clear that the methods developed in this paper can be applied
to other processes as well, both the large $N$ method and
the leading logarithms to higher loop orders for general $N$.
Work is in progress for the decay constant, $\pi\pi$-scattering
and the formfactors \cite{BCwork}.

\section*{Acknowledgments}

This work is supported by the 
Marie Curie Early Stage Training program “HEP-EST” (contract number
MEST-CT-2005-019626),
European Commission RTN network,
Contract MRTN-CT-2006-035482  (FLAVIAnet), 
European Community-Research Infrastructure
Integrating Activity
``Study of Strongly Interacting Matter'' (HadronPhysics2, Grant Agreement
n. 227431)
and the Swedish Research Council.

\appendix

\section{Integrals}
\label{integrals}

We can get rather high powers of momenta in our the integrals. Let us first look
at integrals without external momenta.
These are of the form
\be
\label{tadintegrals}
I_{mn} = \frac{1}{i}\int\frac{d^dq}{(2\pi)^d}\,
\frac{q_{\mu_1}\ldots q_{\mu_m}}{\left(q^2-M^2\right)^n}\,.
\ee
This vanishes for $m$ odd, and for even $m$ we can use\footnote{
The formula is only valid for $d=4$ which is sufficient for
our purpose. We derived it
using recursive methods but it is probably well known
in the higher loop integral community.}
\ba
q_{\mu_1}\ldots q_{\mu_m} &\to& \frac{1}{2^{m/2}\left(\frac{n}{2}+1\right)!}
\left(q^2\right)^{m/2} G_{\mu_1\ldots\mu_m} \,,
\nonumber\\
 G_{\mu_1\ldots\mu_m} &=&
g_{\mu_1\mu_2}\ldots g_{\mu_{m-1}\mu_m}+g_{\mu_1\mu_3}\ldots\,.
\ea
Where $g_{\mu\nu}$ is the metric tensor.
The right hand side of the last term consists of all possible ways to
put the Lorentz indices on the metric tensor $g_{\mu_i\mu_j}$
and is symmetric under all interchanges of the indices.
After that we use recursively
\be
\label{reduceintegrals}
\frac{q^2}{q^2-M^2} = 1+\frac{M^2}{q^2-M^2}
\ee
to obtain terms either without propagators or without powers of $q^2$.
Of the resulting integrals only two have a divergent part
\ba
\frac{1}{i}\int\frac{d^dq}{(2\pi)^d}\,
\frac{1}{\left(q^2-M^2\right)}
&=&\frac{1}{16\pi^2}\,\frac{M^2}{\epsilon}+\mathrm{finite}\,,
\nonumber\\
\frac{1}{i}\int\frac{d^dq}{(2\pi)^d}\,
\frac{1}{\left(q^2-M^2\right)^2}
&=&\frac{1}{16\pi^2}\,\frac{1}{\epsilon}+\mathrm{finite}\,,
\ea
with $d=4-2\epsilon$.

We need to do know the divergent parts of one-loop integrals with up
to 5 propagators for this calculation. This we do by combining propagators
using Feynman parameters and then shifting the momentum variable
to obtain integrals of the type (\ref{tadintegrals}).
The Feynman parameter integrals needed are always simple polynomial ones.

The above procedure can be programmed in FORM to work recursively.

\end{document}